\documentclass[journal ]{new-aiaa}
\usepackage[utf8]{inputenc}
\usepackage{textcomp}
\usepackage[most]{tcolorbox}
\usepackage{graphicx}
\usepackage{amsmath}
\usepackage[version=4]{mhchem}
\usepackage{siunitx}
\usepackage{longtable,tabularx}
\usepackage{subcaption}
\usepackage{authblk}
\usepackage{epstopdf}
\usepackage{blindtext,graphicx}
\usepackage[absolute]{textpos}
\title{Evaluation of Aerodynamic Characteristics in Oscillatory Coning Using CFD Methods}
\author[1]{Mohamed Sereez}
\author[2]{Mikhail Goman}
\affil[1]{School of Mechanical, Aerospace and Automotive Engineering, Coventry University, UK}
\affil[2]{School of Engineering and Sustainable Development, De Montfort University, UK}
\begin{document}
\begin{textblock}{10}(8,1)
\noindent\scriptsize DOI: 10.13009/EUCASS2022-7346 \\
\end{textblock}
\begin{textblock}{15}(5,1.5)
\noindent\scriptsize  9th EUROPEAN CONFERENCE FOR AERONAUTICS AND SPACE SCIENCES (EUCASS) \\
\end{textblock}
\maketitle

\section*{Abstract}

The wind tunnel rotary-balance testing is widely used in aircraft dynamics to characterise aerodynamics at moderate and high angles of attack during stall and spin regimes. In such experiments an aircraft test model is rotated along the wind-tunnel free-stream velocity vector allowing the measurement of aerodynamic characteristics in steady rotational flow conditions with constant angle of attack and sideslip. In modified tests named as oscillatory coning, the rotation vector is tilted from the free-stream velocity vector making flow conditions with periodic variations in angle of attack and sideslip. This allows evaluation of unsteady aerodynamic responses superimposed on steady conical rotation. The use of CFD methods for prediction of aerodynamic characteristics in rotary-balance and oscillatory coning conditions may significantly complement experimental data via extrapolation of data for higher Reynolds numbers, elimination of interference effects from supporting system, extraction of unsteady aerodynamic derivatives affecting aircraft dynamic stability. This paper presents CFD simulation results obtained in rotary-balance and oscillatory coning motions for the NASA Common Research Model (CRM) in its wing-body configuration at moderate $Re=1\times10^6$, low Mach number $M\mbox{=}0.2$ and the use of the obtained unsteady responses in aerodynamic modelling.

\section{Introduction}
Currently, aerodynamic modeling is attracting more and more attention to analyze the dynamics of an aircraft and simulate flight at high angles of attack, which are characteristic of stall and lateral departure. This is dictated by the need to solve the loss of control in flight (LOC-I) safety problem \cite{JA-SUPRA,LowenBergTTail}. The rotary-balance test method was developed to provide information on the effect of angular rotation on the overall aerodynamic forces and moments acting on an aircraft in flight \cite{AGARD-AR-265}. This technique is more suitable for the developed spin modes, but it is also  meaningful for lateral departures in the stall zone. Unsteady aerodynamic effects in a pure conical motion of an aircraft can be additionally evaluated under the conditions of the  oscillatory coning, which occurs when the axis of rotation deviates from the direction of the incoming flow \cite{AGARD-AR-265} (ch. 4 on p. 69). Such deflection results in periodic changes in the angle of attack and side slip on the aircraft model being tested, creating additional non-stationary aerodynamic responses superimposed on aerodynamic response from conical motion.

This paper presents computational predictions of aerodynamic characteristics of the NASA Common Research Model (CRM) \cite{NASACRM} configuration using the CFD simulations. The use of computational methods along with experimental data allows CFD results to be verified, interference effects to be removed from experimental data, results to be extended to higher Reynolds numbers, and more. The open source CFD software OpenFOAM, based on the the finite volume method \cite{OpenFOAMmain, OpenFOAM_4_CFD, MSNBMG} is used to simulate the rotary-balance and oscillatory coning types of motion. The hybrid Reynolds Averaged Navier-Stokes Equations along with Large Eddy Simulations (RANS/LES) is likely to be the most appropriate approach for the prediction of separated flow regions in the post-stall zone, however, in this paper the Unsteady Reynolds Averaged Navier Stokes Equations (URANS) equations along with the Shear-Stress Transport (SST) turbulence model \cite{MenterSST} are used primarily due to two reasons a) the unified grid generation guidelines of the CRM  was not intended for a hybrid RANS/LES simulation \cite{UnifiedCRM,DragPW4} and b) due to the high computational cost of the hybrid simulations. This approach is commonly adopted in aviation applications \cite{Neil,Our1,MSNBMG}. In addition to the dynamic meshing techniques available in OpenFOAM, a new dynamic mesh motion function termed "conical motion" is implemented. Through this mesh motion function the rotation of aircraft around any arbitrary axis at the desired non-dimensional rotation rate can be defined.

In the conical motion, the body-fixed grid rotates around the prescribed velocity vector $\vec V$ and there are no changes in angle of attack and sideslip angle. In the oscillatory coning the body-fixed grid rotates around the the prescribed rotation vector $\vec\Omega$ with a fixed angle of offset $\delta$  from the velocity vector $\vec V$. 

Within the framework of OpenFOAM, both of these dynamic mesh motion functions are carried out by transforming the grid at every time step based on quaternions, allowing a conical rotation of the aircraft around the body-fixed frame of reference $OX, OY, OZ$, in which the body axis projections of aerodynamic forces and moments can be calculated. Implementation of such a force and moment extraction function in OpenFOAM, lead to the elimination of the original calculation of the aerodynamic forces and moments in the wind-axis frame of reference. The rotation of grid cells causes an induced velocity on the boundary surface, which then requires special attention to ensure the no-slip boundary condition on the surface of the aircraft. This type of boundary condition in OpenFOAM is categorized as "movingWallVelocity" boundary condition which ensures that the normal velocity flux across the boundary surface is zero.

In this paper, we present the methodology and capabilities of the open source CFD software OpenFOAM \cite{OpenFOAM,OpenFOAM_4_CFD} to simulate rotary-balance and oscillatory coning testing of aerodynamic characteristics in the post stall zone. The paper is organized as follows; Section "Computational framework" presents the governing equations, grid generation methodology, adopted numerical framework and the kinematics of the rotary-balance and oscillatory coning. Section "Results and Discussions" presents the validation of the OpenFOAM computational results for the NASA CRM model in steady conical motion along with the obtained simulation results for the rotary-balance tests and oscillatory coning unsteady aerodynamic responses. The concluding comments and remarks are outlined in the last section.

\section{Computational framework}

\subsection{Geometry and Grid generation}

The NASA Common Research Model (CRM) is a representative of a commercial transport aircraft with it's cruise Mach number, $M\mbox{=}0.85$ and design lift coefficient $C_L\mbox{=}0.5$ \cite{NASACRM,DragPW4}. The CRM model was specifically designed for the purpose of validation and comparison of computational simulation data amongst different CFD codes and against wind tunnel test results, thus enabling a more unified validation and verification approach for external aerodynamics applications.

\begin{figure}[htb!]
		\centering
		\includegraphics[width=0.7\textwidth]{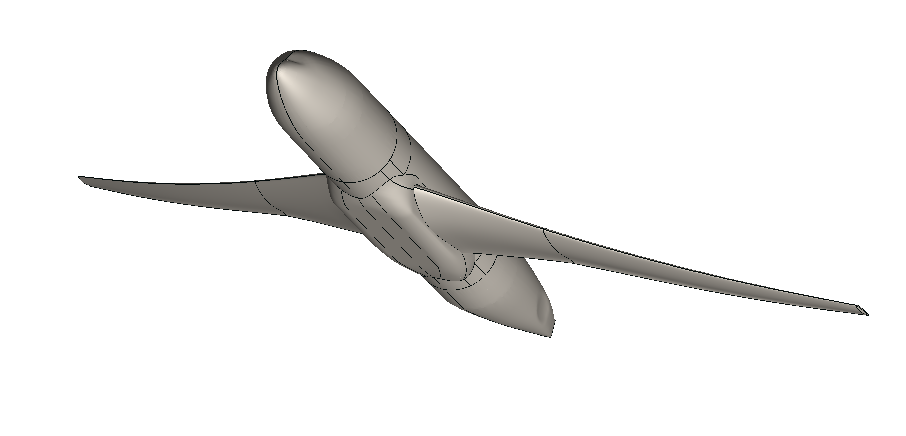}
		\caption{CRM Wing-Body configuration}
		\label{Fig-CRM-geom}
	\end{figure}

The CRM wing is made up of a thin super-critical airfoil  with aspect ratio $AR\mbox{=}9$ and a taper ratio of $0.25$. The CRM model is provided by the drag prediction workshop \cite{DragPW4} in three different configurations a) Wing-Body (WB), b) Wing-Body-Horizontal Tail (WB-HT) and c) Wing-Body-Horizontal Tail-Nacelle-Pylon (WB-HT-NP). The high-lift configuration of the CRM model is also available in \cite{HLPW}. For the purposes of this paper, the Wing-Body (CRM-WB) configuration of CRM aircraft model as shown in Fig. \ref{Fig-CRM-geom} is sufficient. For future reference, instances of "CRM" appearing in this paper refers to the CRM-Wing-Body (CRM-WB) configuration. This model has a reference mean aerodynamic chord of $c_{ref}=7.0m$, full-wing span, $b=58.76m$ and the reference area for the full model is $S_{ref}=383.65m^2$. The model dimensions are further outlined in Table. \ref{Tab:Table1}.

\begin{table}[htb!]
\begin{center}
\captionof{table}{Reference data for the CRM wing-body model (full model)  \label{Tab:Table1}}
\begin{tabular}{cc}
\hline
Wing span, $b$ & $58.76 m$ \\
Mean Aerodynamic Chord, $MAC$, $c_{ref}$ & $7 m$ \\
Reference area  $S_{ref}$ & $383.68 m^2$ \\
Wing aspect ratio, $AR$ & $9.0$ \\
Moment reference point $C_g (m)$ & $X\mbox{=}33.67, Y\mbox{=}0, Z\mbox{=}4.52 $ \\
\hline
\end{tabular}
\end{center}
\end{table}

The computational grids for the CRM model is made by following the guidelines provided by the drag prediction workshop committee \cite{UnifiedCRM}. Recent studies with the CRM model \cite{JA-SUPRA} shows that a "medium" grid size with $10.0\times10^6$ elements is sufficient for the study of flight dynamics applications including estimation of longitudinal and lateral derivatives.

\begin{figure}[htb!]
		\centering
			\fbox{
		\includegraphics[width=0.75\textwidth]{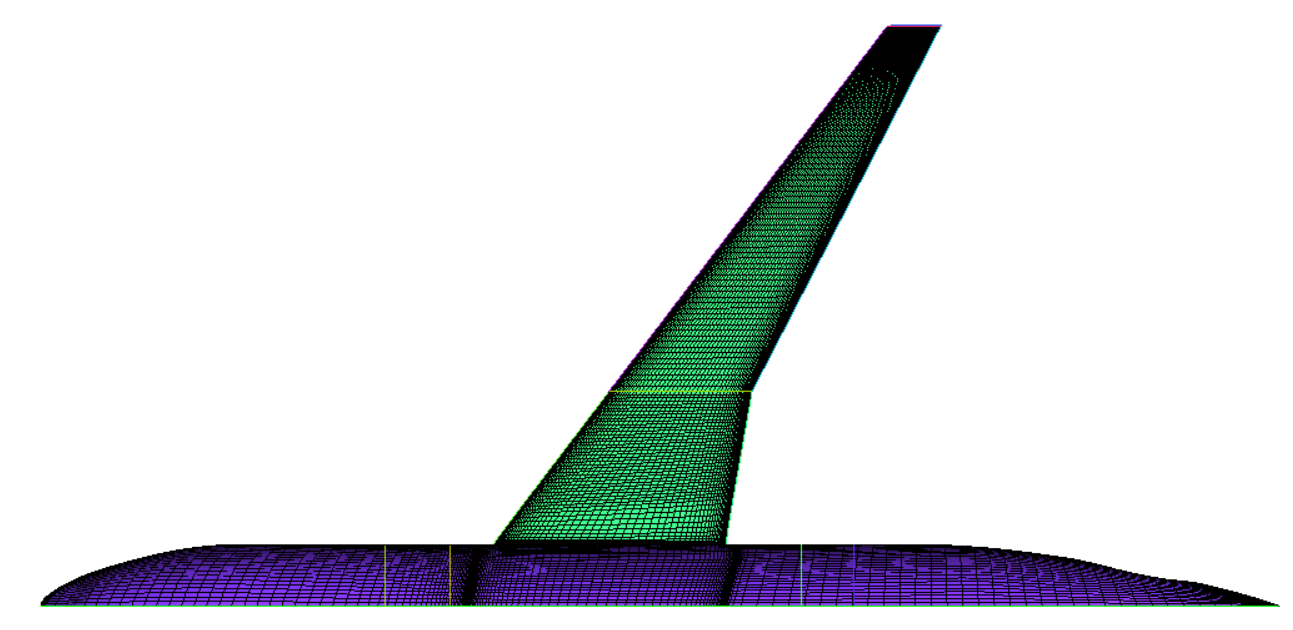}}
		\fbox{
		\includegraphics[width=0.75\textwidth]{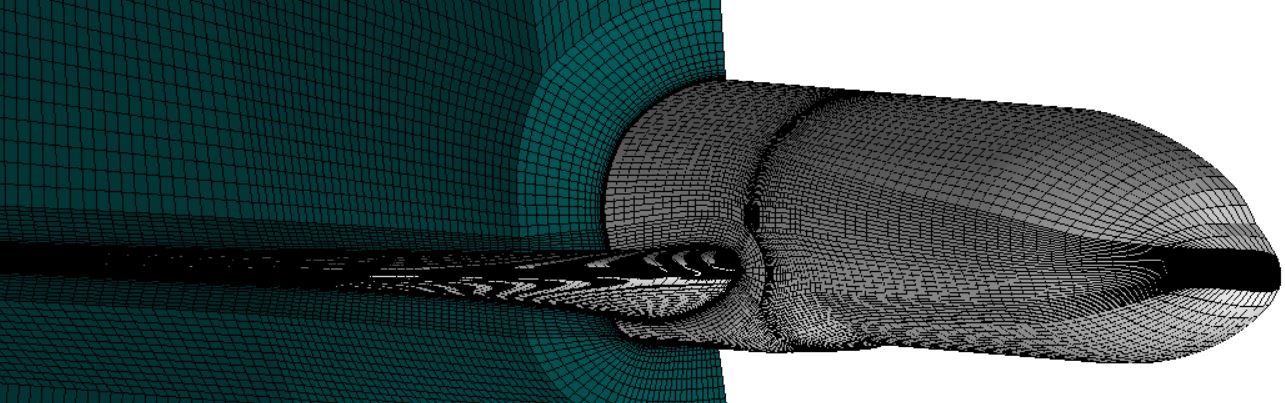}}
		\caption{Adopted grid for the CRM-WB configuration at $Re\mbox{=}1\times10^6$}
		\label{Fig-CRM-grid}
	\end{figure}

The computational grids are made using an H-O grid topology, enabling a smooth wrapping of the O-type blocking around the aircraft model. This allows to generate a high quality structured grid with good cell determinant quality for the surface and volume grid cells. Using an O-type blocking, the boundary layer is well defined and the cell skewness and orthogonality is maintained at the desired optimum values. The cell area and volume transition ratios are in the range of $1.0 - 1.2$ allowing only 20 percent maximum change, thus ensuring no large gradient jumps of the flow scalar and vector variables during the simulation. The variation of cell sizes along the wing are defined as $0.1\%$ of the local span-wise or chord-wise wing length. For the trailing edge a minimum of $5$ cells are placed. The boundary layer consists of $25$ adjacent layers with growth rate of $1.15$.  The first cell layer's height was determined by the non-dimensional wall distance of $Y+\leq1$ enabling to fully resolve the boundary layer. A hyperbolic approach is used to ensure that the first $2\mbox{-}3$ cells of the boundary layer have almost constant height. The resulting grid is shown in \ref{Fig-CRM-grid}. 

For the purpose of consistency, the number of elements in the mesh was maintained as that for the grids used in \cite{JA-SUPRA}, which is 1) coarse grid with 5 million elements and 2) medium grid with 10 million elements. The comparison of the obtained computational results for the steady conical motion of CRM against CFD code ENFLOW in the validation section of this paper shows an accurate match. Furthermore, the use of a non-dimensional physical time step size less than the convective length of the flow, $t=0.1C_{ref}/V$ and force convergence studies shows that the grid size "medium" with $10\times10^6$ elements is sufficient for the purposes of this paper.

\subsection{Governing Equations}
The Navier--Stokes (NS) equations governing incompressible fluid flow are 
the continuity equation 
\begin{equation}
\label{Eq1-Cont}
\nabla \cdot\mathbf{U} = 0
\end{equation}
and the momentum equation
\begin{equation}
\label{Eq2-Momentum}
\frac{\partial \mathbf{U}}{\partial t} + (\mathbf{U}\cdot\nabla)\mathbf{U}  - \nu \nabla^2 \mathbf{U} = -\frac{\nabla \mathbf{p}} {\rho}
\end{equation}		

The computational resources required for Direct Numerical Simulations (DNS) of Eqs.~(\ref{Eq1-Cont}) and (\ref{Eq2-Momentum}), especially for flow conditions with high Reynolds numbers, usually exceed currently available computational capabilities.
Instead the Unsteady Reynolds-Averaged-Navier-Stokes (URANS) equations are solved, in which the Reynolds stresses arising as a result of averaging the fluctuating velocities are described by some additional empirical equations either algebraic or differential to represent an appropriate turbulence model.Most turbulence models for the URANS equations are based on the concept of eddy viscosity, which is equivalent to the kinematic viscosity of a fluid, to describe turbulent mixing or flow momentum diffusion ~ \cite {OpenFOAM_4_CFD}. The Reynolds stresses, which appear in the URANS equations due to time averaging, are described in linear turbulence models with the following Boussinesq assumption:
\begin{equation}
\label{Eq3-Boussinesq}
\tau_{ij} = 2 \nu_t \!\left(S_{ij} -  \frac{1}{3}\frac {\partial u_k}{\partial x_k} \delta_{ij}\right) - \frac{2}{3}\rho k \delta_{ij}1
\end{equation}

\subsection{Numerical framework and boundary conditions}

 For the inlet velocity a Dirichlet boundary condition with fixed velocity vector $V_{inlet}=({V_x, V_y, V_z})$ is applied and for the inlet pressure a Neumann type zero gradient boundary $dp/dx_j=0$ condition is used. The outlet is prescribed a zero gradient velocity $dV/dx_j=0$  and a static pressure of $p_{outlet}=0$. The turbulent kinetic energy $k$ at the inlet is a fixed value estimated using a turbulence intensity value of $0.1\%$,  ensuring replication of common aerodynamic wind tunnel testing conditions. The aircraft also has a "movingWallVelocity" boundary condition to enable zero-flux condition for the dynamic rigid body mesh motion. 
For the evaluation of turbulent viscosity standard SST turbulence model ~\cite{MenterSST} is employed. This approach is commonly used in external aerodynamics involving adverse pressure gradients and strongly separated flow conditions \cite{Russel,Our1}. The two equation $k\mbox{-}\omega$ SST model solves for the prediction of two scalar flow variables which are the turbulent kinetic energy $k$ and specific dissipation rate of turbulence $\omega$.

On testing of various finite volume schemes and solvers of OpenFOAM the Pre-conditioned Conjugate (PCG) solver with Geometric Algebraic Multi-Grid (GAMG) as a pre-conditioner was found out to be the most efficient algorithm. Employing GAMG pre-conditioner with 10-30 iterations and applying pre and post smoothing of the residuals for $2-3$ levels, ensures that only about about $30-50$ iterations of the PCG linear solver is needed to drive the residuals to near zero values at each time step. Additionally, the PCG solver was found out to be more robust for grid size of 10 million elements in the parallel approach of OpenFOAM. The gradients of the flow quantities are quantified using the second order accurate Gauss linear scheme with limiters based on cell center values of the flow quantities. The  divergence of the velocity field and the scalar turbulent quantities are also estimated using second order accuracy with the "cellLimited Gauss linear" scheme of OpenFOAM. For estimation of contribution of cell center variables to the faces a linear interpolation is used.

The computational simulations using OpenFOAM are carried out on the high performance computing cluster  "Zeus" \cite{ZeusHPC} (Heterogeneous HPC cluster with 3200 CPU cores) at Coventry University using 4 compute nodes, 32 CPU-cores per node and a RAM of 128gb per node. For the purpose of fast, robust and accurate unsteady simulations an implicit dual-time stepping method is implemented and used in the  OpenFOAM compiled in this cluster. The dual time framework formulated as in  from\cite{JA-Dual-NASA,StengelDynamic} is commonly adopted for aerodynamic flows with rigid body mesh motion such as  dynamic stall simulations.

The dual time stepping framework can be described as:

\begin{equation}
\label{Eq4-General}
\frac{dW}{dt}+ R(W) = 0
\end{equation}
where W is the flow variables at grid points and R is the residuals of flow variables, including the dissipative terms. The discretized equation \ref{Eq4-General} is then formulated in dual-stepping in conjunction with Euler backward time integration technique as follows:
\begin{equation}
\label{Eq5-dual}
\frac{dW}{dt^*} + \frac{3}{2\Delta t}W - \frac{4}{2 \Delta t}W^n + \frac{1}{2\Delta t}W^{n-1} + R(W) = 0
\end{equation}
 In Eq. \ref{Eq5-dual} $t*$ is the pseudo time step size, $\Delta t$ is the physical time step size and $n$ is the time step number of the simulation.
 
The dual time stepping contribution to an assembled finite volume matrix with $A$ being diagonal coefficients and $H$ being the off diagonal coefficients is as follows. 

\begin{equation}
\label{Eq-set6}
    A +=  \bigg(\frac{3}{2 \Delta t}+ \frac{1}{\alpha_{k} \Delta \tau}\bigg) \int_{S} dSdA
\end{equation}

\begin{equation}
\label{Eq-set7}
   H' += \Bigg(  \bigg(\frac{1}{\alpha_{k} \Delta \tau}\bigg)U^{m-1} + \bigg(\frac{2}{\Delta t}\bigg)U^{n-1}   - \bigg(\frac{1}{2 \Delta t}\bigg)U^{n-2}  \Bigg) \int_{S} dSdA
\end{equation}
where is\textit{n} is the time step level, \textit{m} is the sub-iteration level within each time step, $\alpha_k$ is the Runge-Kutta coefficient in each stage and $ \int_{S} dSdA$ is the integrated volume of each element.

\subsection{Rotary Balance testing setup in OpenFOAM}

The rotational flow and the induced aerodynamic responses to the aircraft model depend on the type of rotational motion.
The differences in the kinematics of rotary-balance conical motion and oscillatory coning motion are described below.

\subsubsection{Rotary-balance: Conical motion setup}

\begin{figure}[htb!]
		\centering
		\includegraphics[width=0.7\textwidth]{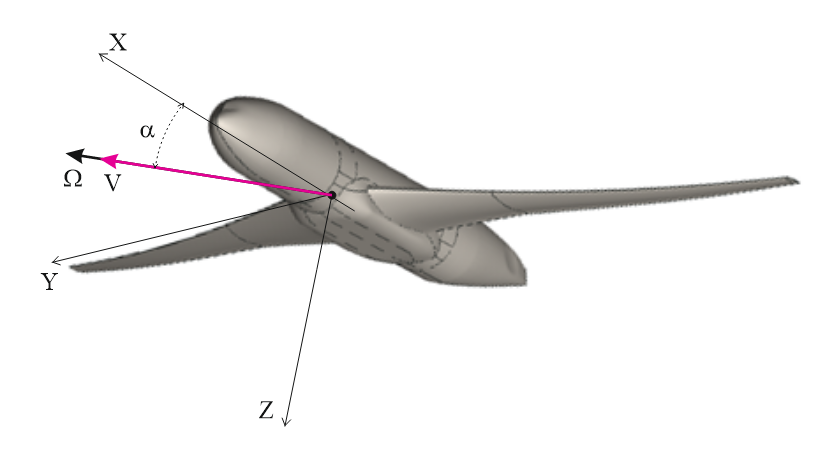}
		\caption{Kinematics of an aircraft model in conical motion - $\alpha=const$, $\beta =const$.}
		\label{Fig-ConicalProjection}
\end{figure}

With a steady conical motion, the aircraft model rotates around the velocity vector  $\vec V$, while the angle of attack $\alpha$, and sideslip angle $\beta$ remain constant throughout the movement, since the rotation vector $\vec\Omega$ coincides with the velocity vector $\vec V$ as shown in Fig\ref{Fig-ConicalProjection}.

%The rotation of the entire aircraft around an arbitrary axis leads to a conically-rotating body axis frame of reference. However, the forces and moments in OpenFOAM are calculated in the fixed inertial reference frame which is the wind-velocity axis. Therefore, it is quite evident that the forces and moments need to be projected back to the body axis frame. In order to achieve this, at the end of every time step and movement of the grid, a quaternion transformation is applied to the  force and moment output. Another approach which can potentially be used to project the forces and moments to the rotating body-axis frame, is to employ the the Euler-Rodriguez transformation. 

The projections of aerodynamic forces and moments in OpenFOAM are calculated in the fixed inertial reference frame directed along the free-stream velocity. Aircraft dynamics require aerodynamic projections on the body-fixed frame of reference $OXYZ$, which rotates with aircraft model. In order to achieve this, at the end of every time step and movement of the grid, a quaternion transformation is applied to the  projections of the aerodynamic forces and moments generated in OpenFOAM. 

The rotary-balance type of conical motion is achieved with new dynamic mesh motion implemented in OpenFOAM termed "conicalMotion" and a sample setup is shown below:

	solidBodyMotionFunction conicalMotion; \\
	origin (33.7 4.52 0 );  // rotation origin \\
	rotationAxis (-0.99756 -0.06976 0);  // rotation axis \\
	rotationOmega 0.23348;  // rotation rate in rad/s \\

\subsubsection{Oscillatory coning motion setup}
\begin{figure}[htb!]
\centering
\includegraphics[width=0.7\textwidth]{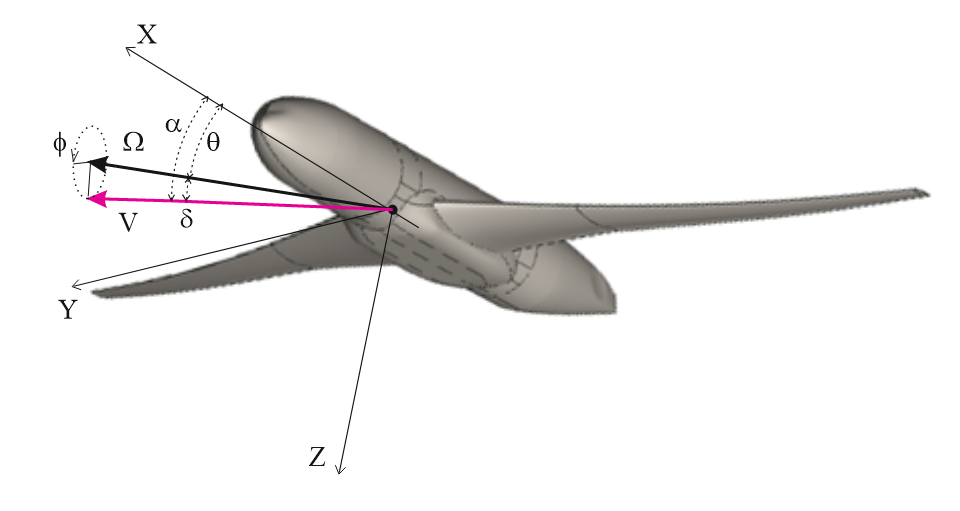}
\caption{Kinematics of an aircraft model in oscillatory conical motion - $\alpha=\theta+\delta \cos \Omega t$, $\beta = \delta \sin \Omega t$.}
\label{Fig-OscConicalProjection}
\end{figure}

In oscillatory coning motion, the rotation vector $\vec\Omega$ deviates from the velocity vector $\vec V$ by an angle $\delta$, as shown in Fig. \ref{Fig-OscConicalProjection}, for example, with a deviation only in the plane of symmetry. More generally, the rotation vector $\vec\Omega$ can have a lateral projection giving a non-zero sideslip center position. In the oscillatory coning, the kinematics of motion of the aircraft model is characterized by a sinusoidal change in the angle of attack $\alpha(t)$ and the angle of sideslip $\beta(t)$ with amplitudes equal to the angle $\delta$ between the angular vectors $\vec\Omega$ and $\vec V$. For an observer inside a body-fixed reference frame, the vector $\vec V$ rotates around the vector $\vec\Omega$ on a conical surface with an angle at the vertex $\delta$.

The roll angle $\phi$ during conical rotation changes linearly with time in accordance with a fixed rotation rate $\Omega$:
\begin{equation}
\label{Eq-OscCon1}
\phi = \Omega t
\end{equation}

And variations of angles of attack, $\alpha$ and sideslip, $\beta$ can be approximately represented in the following form:
\begin{equation}
\label{Eq-OscCon2}
\alpha(t)=\theta+\delta \cos \phi; \,\,\, \beta(t)=\delta \sin \phi
\end{equation}

\begin{figure}[htb!]
\centering
\includegraphics[width=0.49\textwidth]{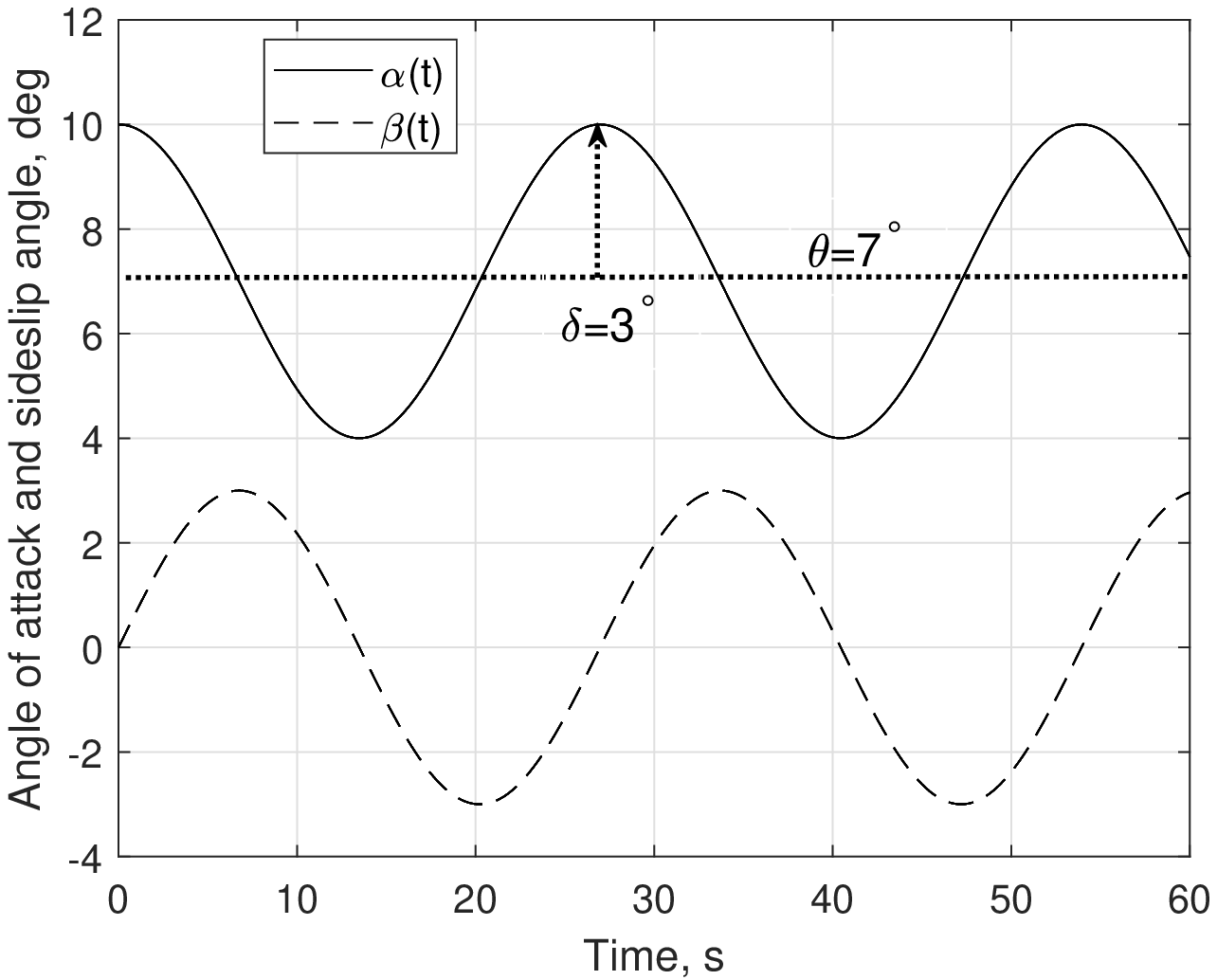}
\includegraphics[width=0.49\textwidth]{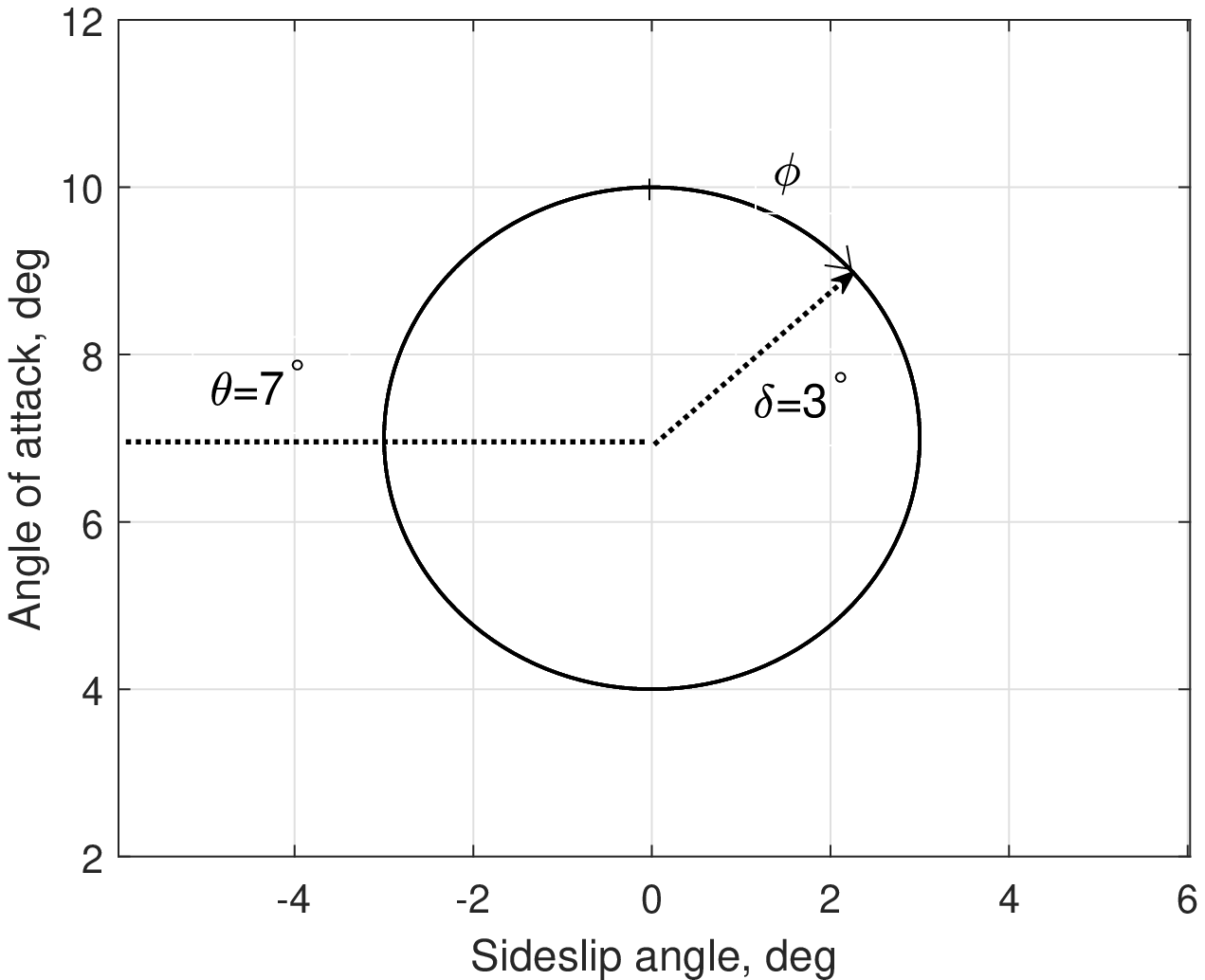}
\caption{Kinematics of oscillatory coning at $\theta=7^\circ$,  $\delta=3^\circ$, $\omega=\Omega b/2V=0.1$.}
\label{Fig-VariationAlfaBeta}
\end{figure}

Time variations of the angle of attack $\alpha$ and sideslip $\beta$ are shown for the installation angle $\theta=7^\circ$ and the deflection angle $\delta=3^\circ$ in Fig.\ref{Fig-VariationAlfaBeta} (left). The oscillatory coning kinematics may be also expressed as a uniform movement along a circular trajectory in the plane of angle of attack and sideslip with a radius equal to the offset angle $\delta$ and a center defined by offset angle $\theta$ as shown in Fig. \ref{Fig-VariationAlfaBeta} (right). 

\section{Simulation results and discussion}

\subsection{Results of rotary-balance for CRM wing-body configuration}

\begin{figure}[htb!]
\centering
\includegraphics[width=0.7\textwidth]{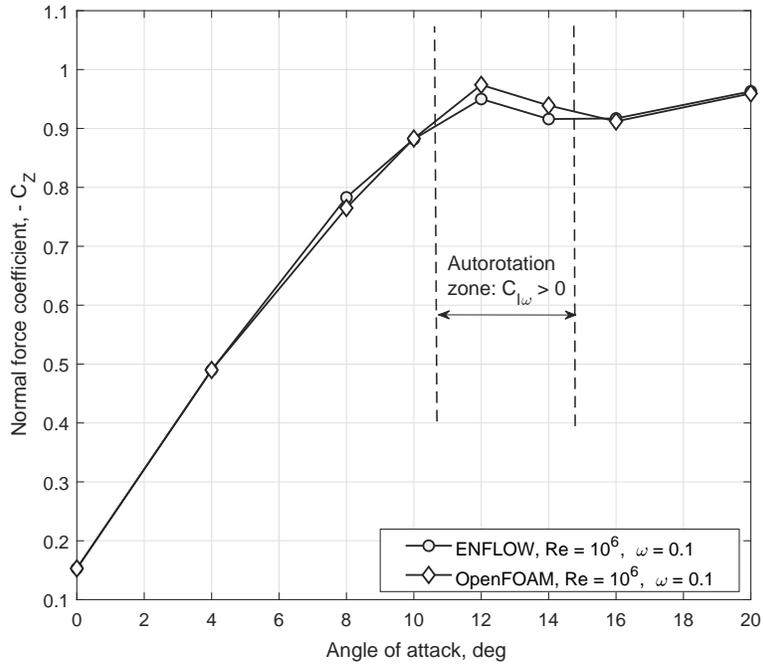}
\caption{Computational prediction of the normal force coefficient $C_Z$ during rotary-balance movement at $\beta = 0$, $\omega=0.1$ for CRM wing-body configuration.}
\label{Fig-CRMValidateCz}
\end{figure}

\begin{figure}[htb!]
\centering
\includegraphics[width=0.7\textwidth]{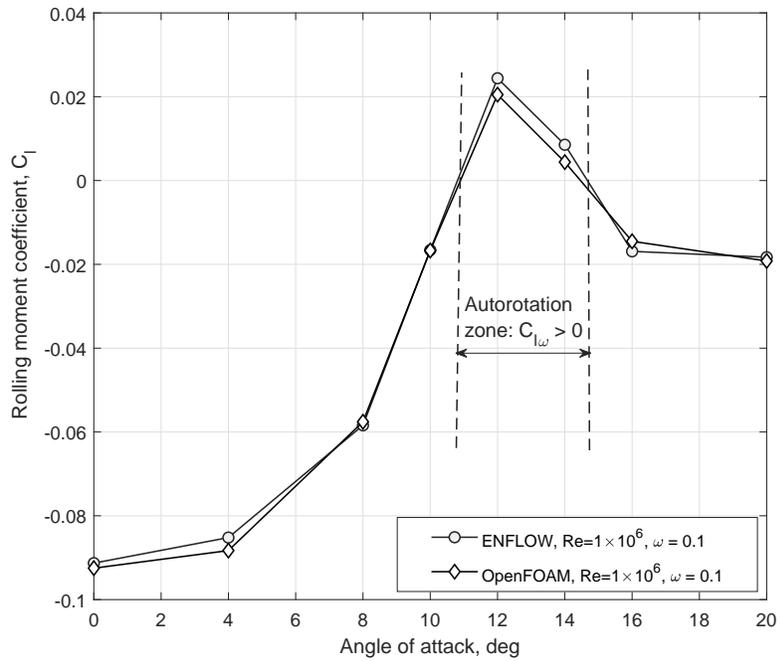}
\caption{Computational prediction of the rolling moment coefficient $C_l$ during rotary-balance movement at $\beta = 0$, $\omega=0.1$ for CRM wing-body configuration.}
\label{Fig-CRMValidateCl}
\end{figure}

Validation of the formulated computational framework of OpenFOAM for the rotary-balance motion was carried out by comparison with the CFD simulation results for the CRM wing-body configuration using the NLR ENFLOW CFD codes presented in \cite{JA-SUPRA}.
The results of the rotary-balance motion simulation in OpenFOAM generally agree well with the results of the ENFLOW CFD code. This is clearly seen in Fig. \ref{Fig-CRMValidateCz} for the normal force coefficient $C_Z$ and in Fig.\ref{Fig-CRMValidateCl} for the rolling moment coefficient $C_l$ calculated at non-dimensional rotation rate $\omega = 0.1$.

There is generally a rather good correlation between the results from \cite{JA-SUPRA} and OpenFOAM both at low angles of attack and in the stall region at high angles of attack with developed separation zones.  The roll damping at low angles of attack, defined by aerodynamic derivative $C_{l\omega} < 0$, degrades with increase of angle of attack due to propagation of separation along the wing. In the stall region this derivative changes its sign $C_{l\omega} > 0$ indicating onset of autorotation regime in the region with negative slope in the normal force coefficient $C_{Z\alpha} < 0$. The minor differences in simulation results of the NLR's ENFLOW and OpenFOAM may be addressed to the employed numerical schemes for the evaluation of diffusive and advective fluxes along with variation in the adopted gradient calculation schemes. 

The conducted validation study strengthens our confidence in the implemented procedure of the rotary-balance movement and transformation of aerodynamic forces and moments into a frame of reference fixed with the body in OpenFOAM.

\begin{figure}[htb!]
\centering
\includegraphics[width=0.75\textwidth]{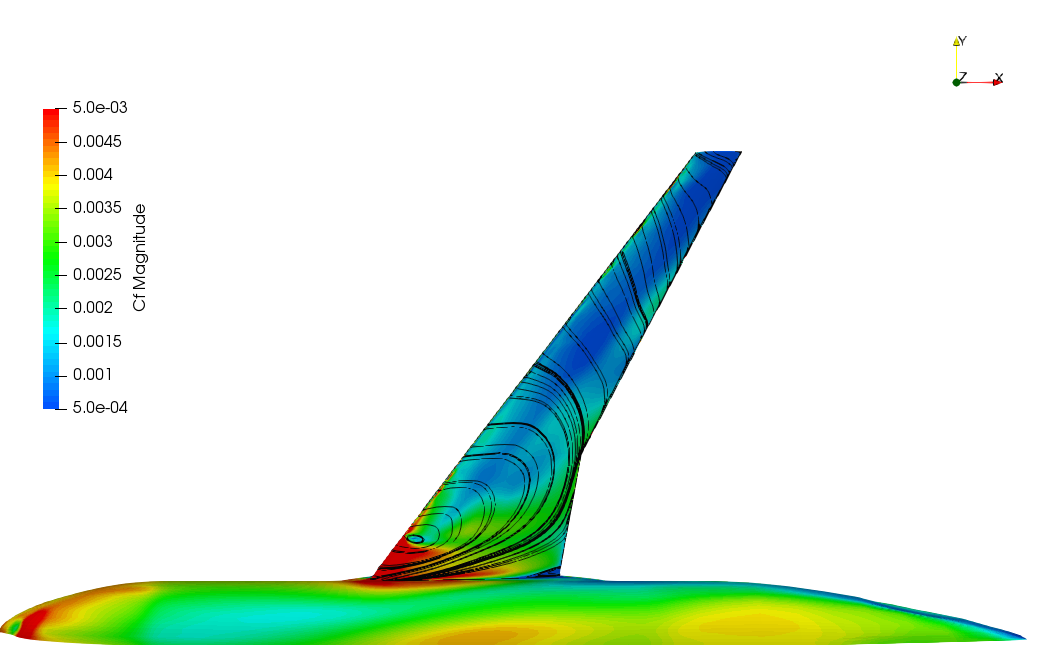}
\caption{Skin friction coefficient $C_f$ for the CRM model at $\alpha = 14^\circ$, $\omega=0.1$ with flow conditions $Re = 1\times10^6$, $M = 0.2$.}
\label{Fig-VisualCRM-Cf-A14Steady}
\end{figure}

\begin{figure}[htb!]
\centering
\includegraphics[width=0.48\textwidth]{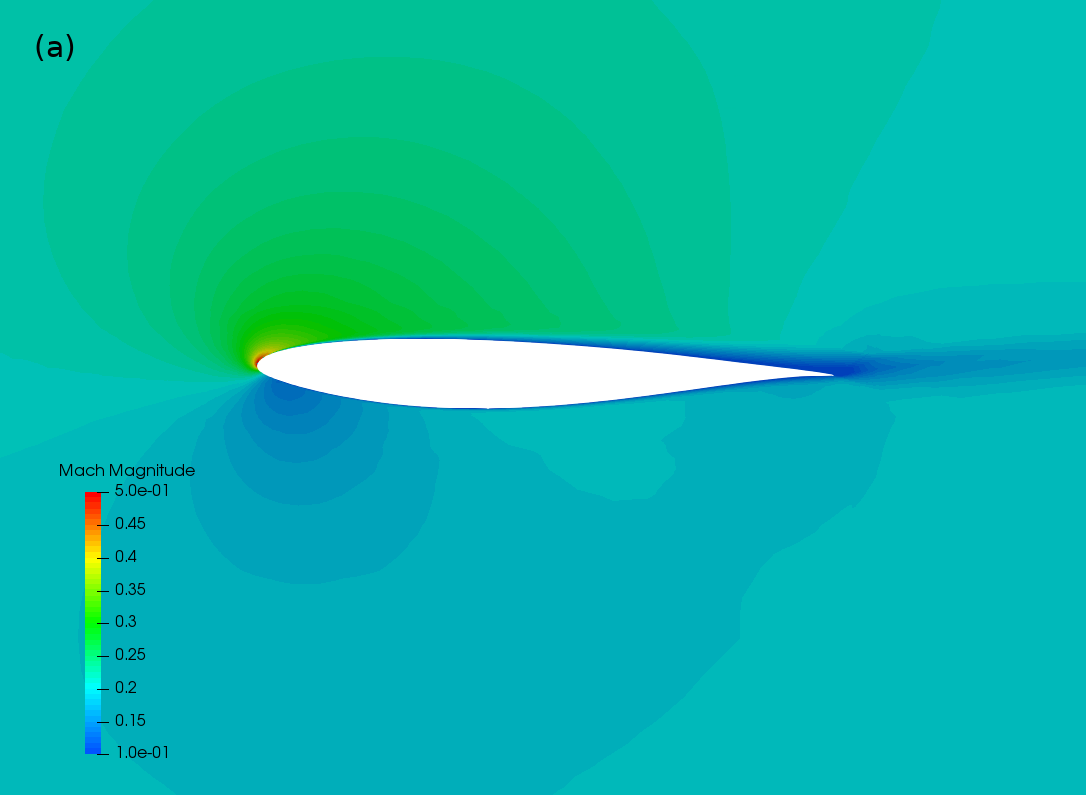}
\includegraphics[width=0.48\textwidth]{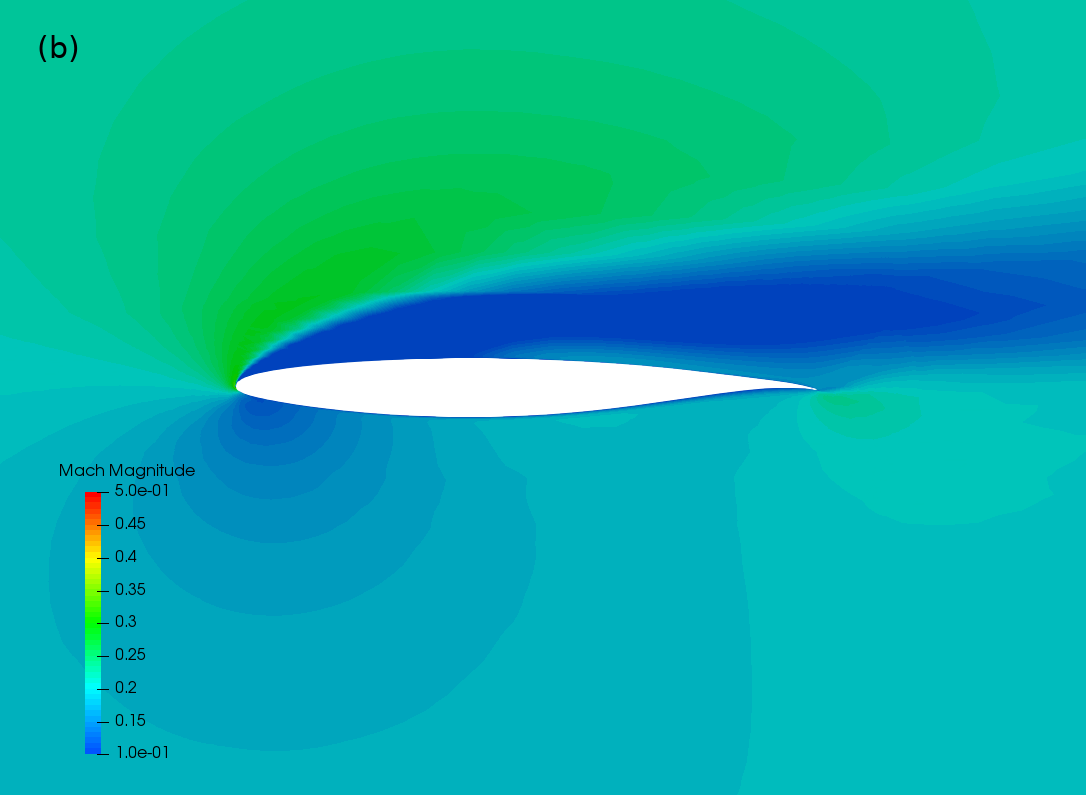}
\caption{Contours of Mach number for the left (a) and right (b) wing in vertical cross-sections placed at $Y=\pm 0.37 \, b/2$ in rotary-balance motion with $\omega = 0.1$ at $\alpha = 14^\circ$ $Re = 1\times10^6$, $M = 0.2$.}
\label{Fig-CompareStarBoardAndPortSide}
\end{figure}	

Visualization of the flow parameters during rotary-balance motion on the CRM surface and around the wing is shown in Fig. \ref{Fig-VisualCRM-Cf-A14Steady} and Fig. \ref{Fig-CompareStarBoardAndPortSide}.
The streamlines with contours of the skin friction coefficient $C_f$ are shown in Fig. \ref{Fig-VisualCRM-Cf-A14Steady} for the right wing upper surface during the rotary-balance motion with $\alpha=14^\circ$ and $\omega=0.1$. This shows a full flow separation across the wing defined by the area with low skin friction. At the same time the flow on the left wing is practically fully attached. 

This statement is supported by the visualized contours of the Mach number placed at wing vertical cross-sections with coordinates $Y=\pm 0.37\, b/2$ shown in Fig. \ref{Fig-CompareStarBoardAndPortSide}. It is evident that on the left wing the flow is fully attached while the right wing the flow is massively separated with significant loss in the normal force leading to the auto-rotation rolling moment. As indicated in Fig. \ref{Fig-CRMValidateCl} the rolling moment coefficient at $\alpha=14^\circ$ is positive $C_l  > 0$ at $\omega = 0.1$ demonstrating that the aircraft is in the auto-rotation regime.

\subsection{Results for oscillatory coning motion}

In an oscillatory coning motion, the angle of attack and sideslip periodically change, causing variations in aerodynamic responses. Aerodynamic variations can also be expected to have periodic variations, but not necessarily single harmonics such as periodic changes in angle of attack and sideslip. After running the simulation, it takes some time for the aerodynamic characteristics to converge to ideal periodic processes, and this initial transient must be excluded from further analysis.

\begin{figure}[htb!]
\centering
\includegraphics[width=0.8\textwidth]{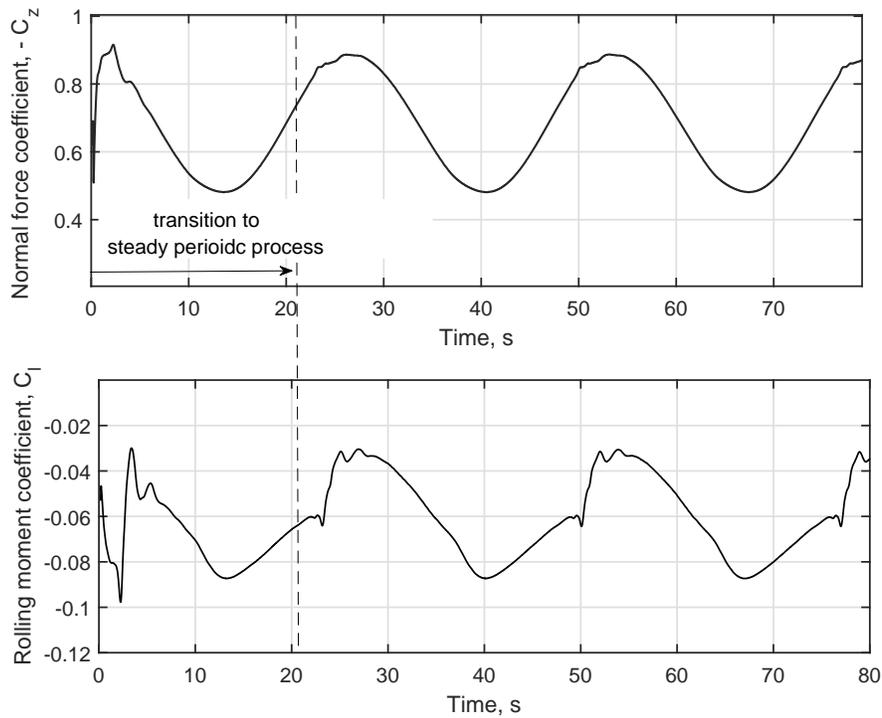}
\caption{Convergence of aerodynamic coefficients $C_Z$ and $C_l$ in oscillatory coning with $\theta\mbox{=}7^\circ$, $\delta\mbox{=}3^\circ$, and $\omega = 0.1$}
\label{Fig-ConvOscillatoryConing}
\end{figure}

The process of convergence of the normal force coefficient $C_Z$ and the rolling moment coefficient $C_l$ is shown in Fig. \ref{Fig-ConvOscillatoryConing} for parameters $\theta=7^\circ$, $\delta=3^\circ$ and $\omega = 0.1$. It can be seen that after the physical time $t\geq20s$ the changes in both aerodynamic coefficients become ideally periodic. The normal force coefficient $C_Z$ is close to the carrier harmonic, but the rolling moment coefficient $C_l$ includes segments with high harmonics due to higher sensitivity to flow separation processes.

\begin{figure}[h!]
\centering
\includegraphics[width=0.72\textwidth]{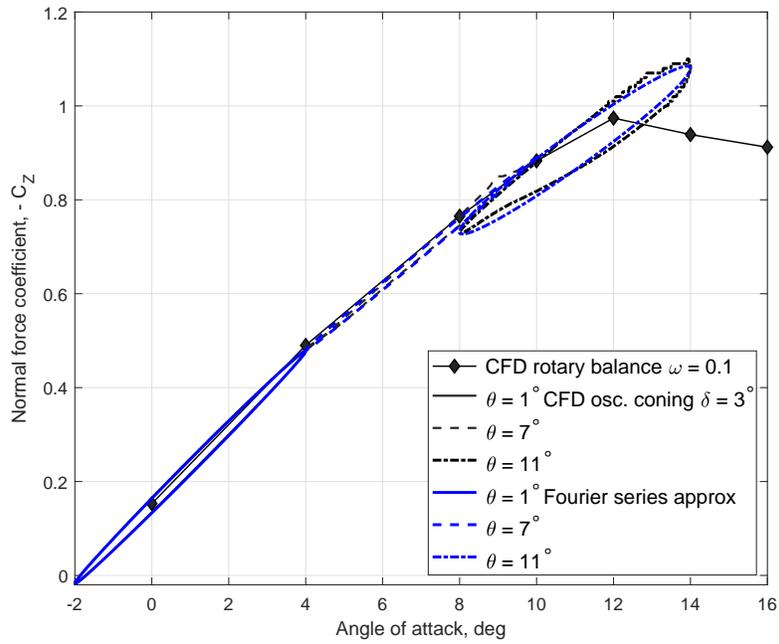}
\caption{Computational prediction of the normal force coefficient $C_Z$ during oscillatory conical motion at $\omega = 0.1$, $M=0.2$ and $Re=1.0\times10^6$.}
\label{Fig-OscCFDresultsCZ}
\end{figure}

\begin{figure}[h!]
\centering
\includegraphics[width=0.72\textwidth]{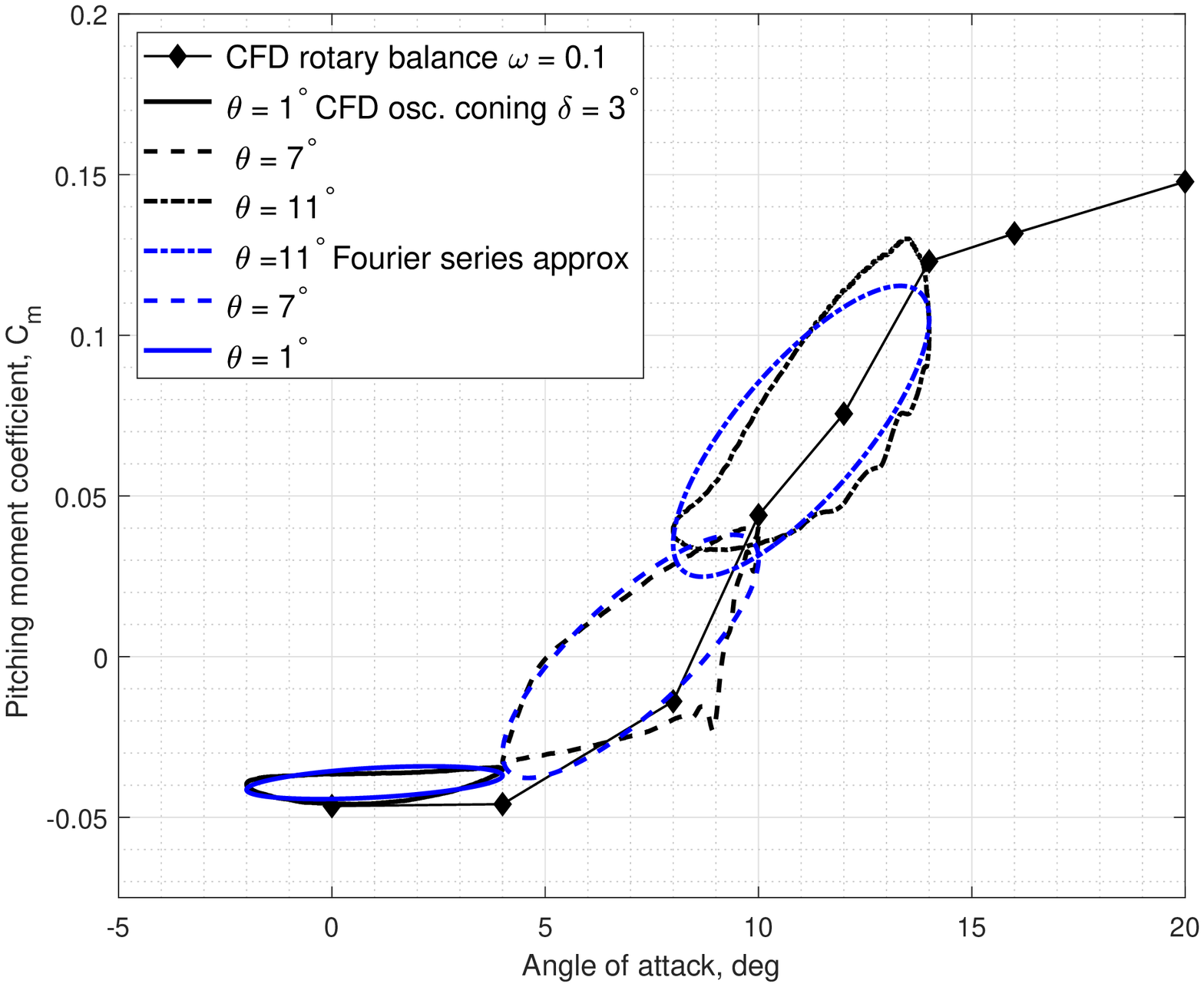}
\caption{Computational prediction of the pitching moment coefficient $C_m$ during oscillatory conical motion at $\omega = 0.1$, $M=0.2$ and $Re=1.0\times10^6$.}
\label{Fig-OscCFDresultsCm}
\end{figure}

\begin{figure}[h!]
\centering
\includegraphics[width=0.72\textwidth]{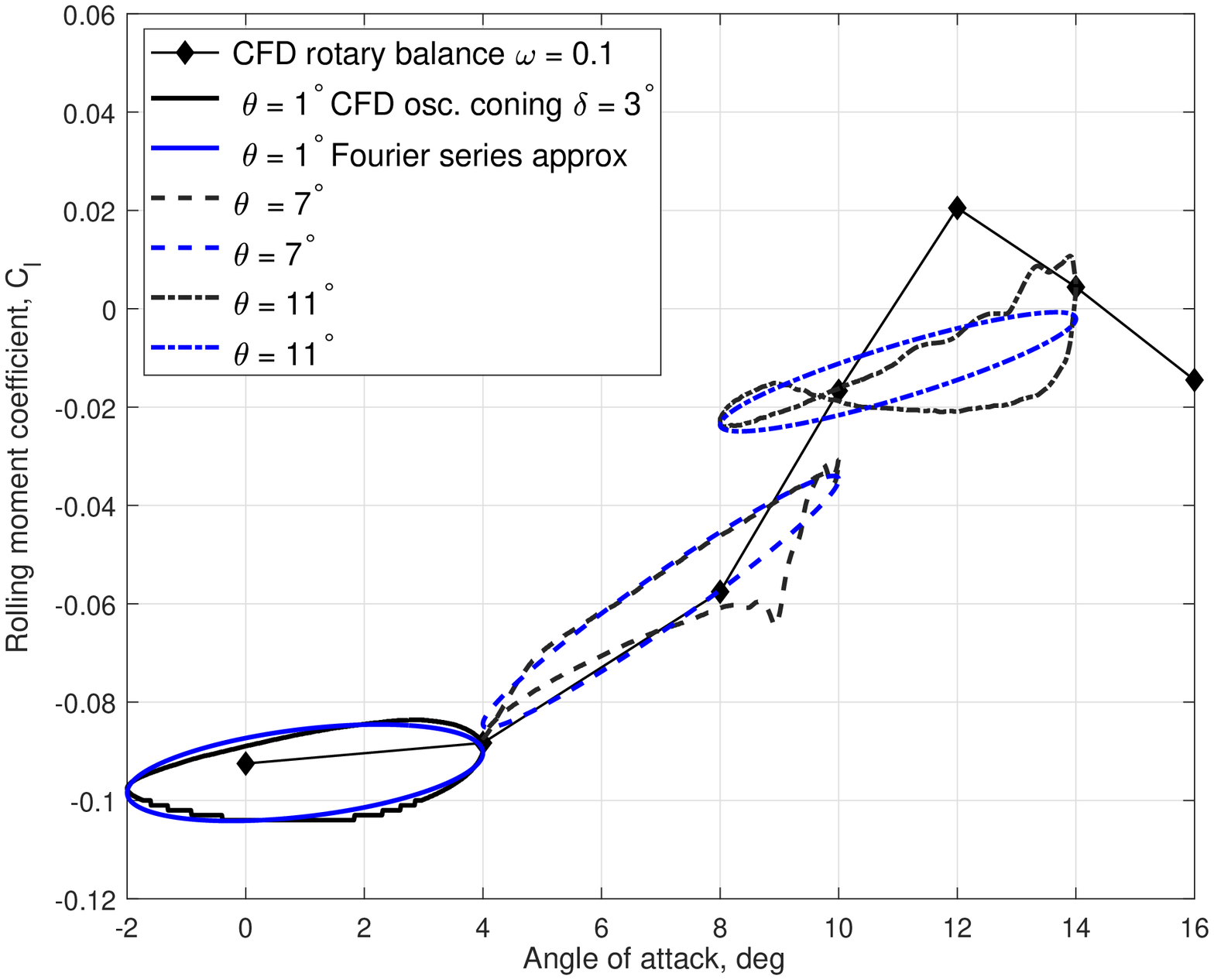}
\caption{Computational prediction of the rolling moment coefficient $C_l$ during oscillatory conical motion at $\omega = 0.1$, $M=0.2$ and $Re=1.0\times10^6$.}
\label{Fig-OscCFDresultsCl}
\end{figure}

%%%%
The OpenFOAM simulation results obtained in oscillatory coning motions for three settings in rotation axis with $\theta = 1^\circ, 7^\circ, 11^\circ$ and offset angle $\delta = 3^\circ$ with non-dimensional rotation rate of $\omega = 0.1$ are shown in Figs. \ref{Fig-OscCFDresultsCZ} - \ref{Fig-OscCFDresultsCl}.
The simulations were conducted at flow conditions with $Re = 1\times10^6$, $M=0.2$.

Non-stationary aerodynamic reactions in the oscillatory coning at different installation angles $\theta$ and displacement angle $\delta=3^\circ$ in the form of closed trajectories are imposed on the static dependencies on the angle of attack obtained in conical motions of rotary-balance tests.

The geometric shapes of non-stationary aerodynamic loops reflect the possibility of their representation in a linearized form. If the non-stationary response loops are close to an elliptical shape and fit well around the static curve, this indicates that the aerodynamic response can be well approximated using the static dependence plus the non-stationary aerodynamic contribution proportional to the rates of change of angle of attack and side slip in the following form ($i = Z, m, l$) \cite{JA-SUPRA}:
\begin{equation}
\label{LinearUnsteadyResponse}
C_i = C_{i_{rb}}(\alpha,\beta,\omega)+C_{i\dot\alpha_{oc}} \dfrac{\dot\alpha c}{2V}+C_{i\dot\beta_{oc}} \dfrac{\dot\beta b}{2V}
\end{equation}
where $C_{i_{rb}}(\alpha,\beta,\omega)$ is steady aerodynamic components identified in the rotary-balance tests, $C_{i\dot\alpha_{oc}}$ and $C_{i\dot\beta_{oc}}$ are unsteady aerodynamic derivatives from oscillatory coning aerodynamic responses. To extract aerodynamic derivatives $C_{i\dot\alpha_{oc}}$ and $C_{i\dot\beta_{oc}}$ the periodic aerodynamic responses $C_i(t)=C_i(t+2\pi/\Omega)$ measured in oscillatory coning motions should be approximated by the first three terms of the Fourier series expansion:
\begin{equation}
\label{FourierExpansion}
C_i(t)=C_{i0}+C_{is}\sin(\Omega t)+C_{ic}\cos(\Omega t)
\end{equation}
where
\begin{equation}
\label{AeroDeriv}
\begin{array}{l}
C_{i0}=C_{i_{rb}}(\alpha_0, \beta_0, \omega)\\
C_{is}=\left (C_{i\beta_{rb}} - \dfrac{\omega c}{b} C_{i\dot\alpha_{oc}}\right )\delta\\
C_{ic}=\left (C_{i\alpha_{rb}} + \omega C_{i\dot\beta_{oc}}\right )\delta\\
\end{array}
\end{equation}

The Fourier approximation of the closed loop unsteady responses in Figs. \ref{Fig-OscCFDresultsCZ} - \ref{Fig-OscCFDresultsCl} for the normal force, pitching and rolling moment coefficients respectively are shown by blue ellipses defined by coefficients $C_{i0}$, $C_{is}$ and $C_{ic}$ from the Fourier series approximation (\ref{FourierExpansion}). One can see that linear approximations of aerodynamic responses in the form (\ref{LinearUnsteadyResponse}) are mostly acceptable except of aerodynamic responses in the normal force and the rolling moment coefficients covering the stall zone at setting angle $\theta = 11^\circ$.

\begin{figure}[h!]
\centering
\includegraphics[width=0.72\textwidth]{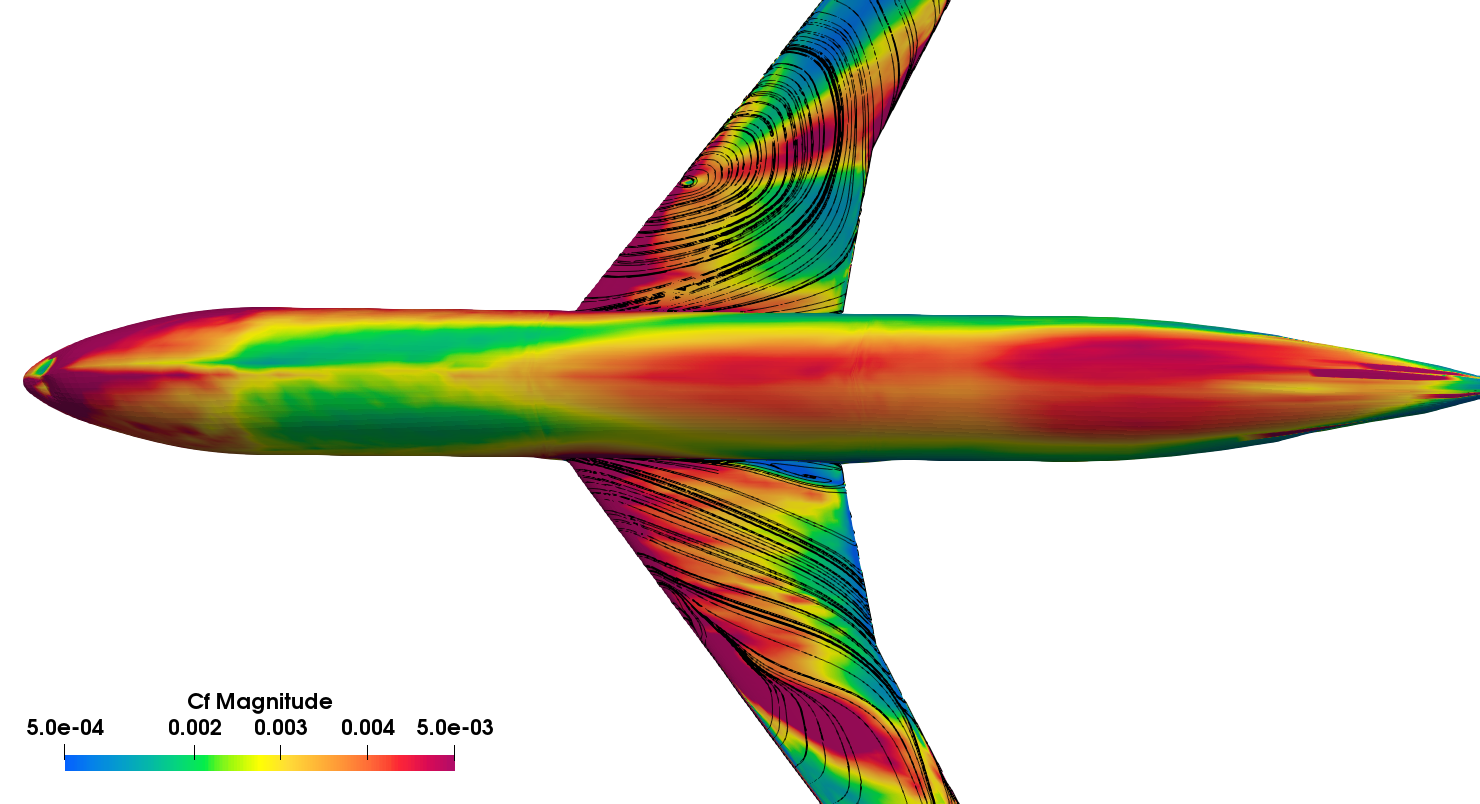}
\caption{Surface streamlines superimposed on contours of skin friction coefficient $C_f$ during oscillatory conical motion at $\alpha(t)=13.3^\circ$, $\beta=1.96^\circ$ and $\omega \mbox{=}0.1$}
\label{Fig-VisualCf-ConicalOsillatory}
\end{figure}

\begin{figure}[h!]
\centering
\includegraphics[width=0.72\textwidth]{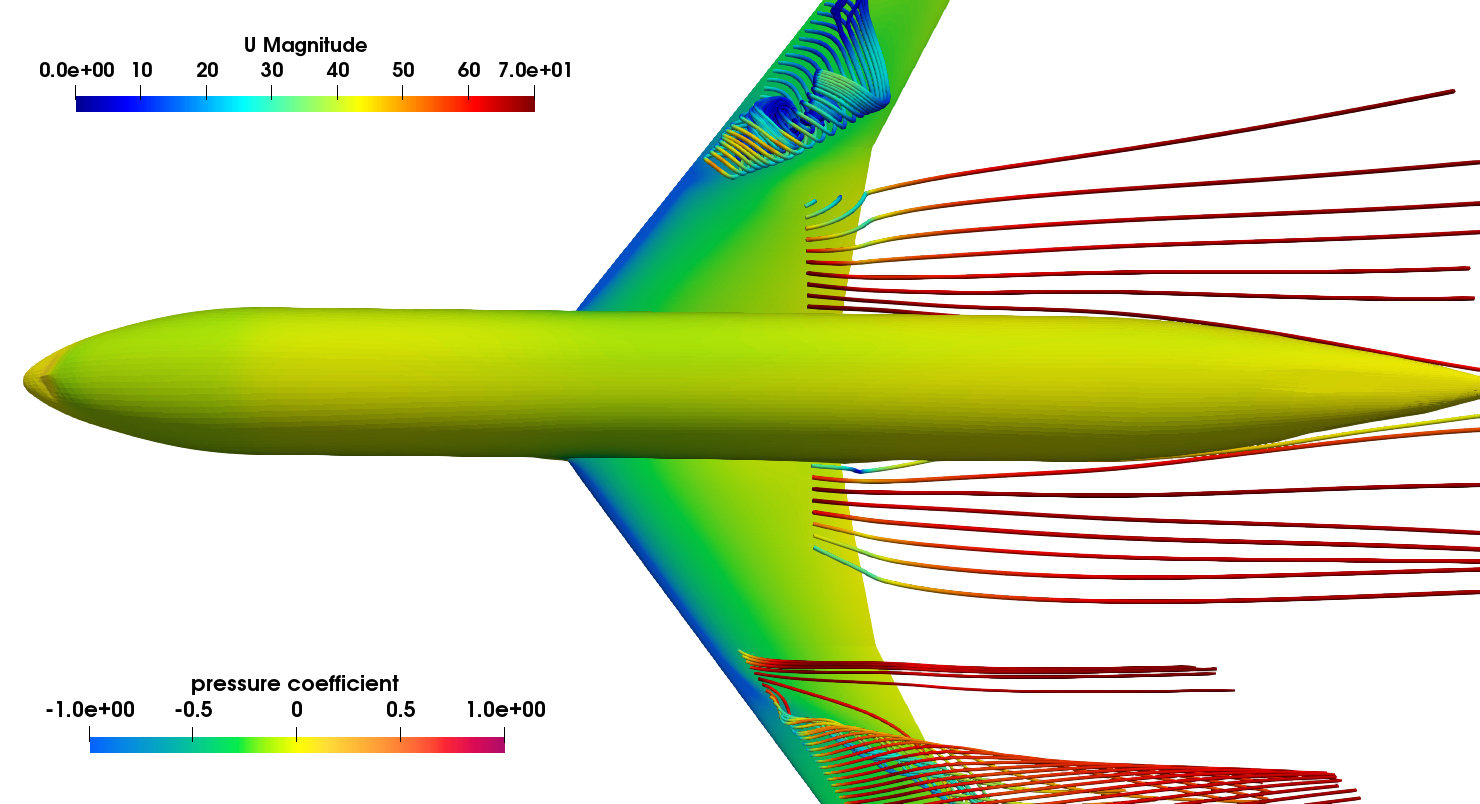}
\caption{Contours of pressure coefficient $C_p$ on aircraft surface and three-dimensional streamlines of velocity during oscillatory conical motion at $\alpha(t)=13.3^\circ$, $\beta=1.96^\circ$ and $\omega \mbox{=}0.1$}
\label{Fig-Visual3d-ConicalOsillatory}
\end{figure}

Visualization of flow parameters during oscillatory coning motion at different times can provide important information for the interpretation of non-stationary aerodynamic characteristics and help in the development of phenomenological aerodynamic models in critical flight conditions. Examples of such visualization are shown in Fig. \ref{Fig-VisualCf-ConicalOsillatory} and Fig. \ref{Fig-Visual3d-ConicalOsillatory}.

\section{Conclusions}

Within the framework of the open source CFD code OpenFOAM, a computational approach to modeling aerodynamic characteristics during rotary-balance and oscillatory coning motions is presented.
Verification of the results of the OpenFOAM rotary-balance simulation showed very good agreement with the results obtained using the CFD ENFLOW code developed at the Royal Netherlands Aerospace Center (NLR). The presented new OpenFOAM capabilities for modeling rotary-balance and oscillatory coning aerodynamic characteristics can make an important contribution to adequate aerodynamic modeling in the extended flight envelope of modern transport aircraft.

\bibliographystyle{unsrt}
\bibliography{citations}

\end{document}